\documentclass[prl,twocolumn,nofootinbib,10pt]{revtex4}

\usepackage{graphicx,amssymb,amsmath}

\newcommand{\beq}{\begin{equation}}
\newcommand{\eeq}{\end{equation}}
\newcommand{\eg}{{\it e.g.}}

\def\mysection#1{{{\bf #1}.~}}

\begin{document}

\title{Implications of the Measurement of Ultra-Massive Boosted Jets at CDF}
\author{Yochay Eshel, Oram Gedalia, Gilad Perez and Yotam
Soreq} \affiliation{Department of Particle Physics \&
Astrophysics, Weizmann Institute of Science, Rehovot 76100,
Israel}

\begin{abstract}
The CDF collaboration recently reported an upper limit on
boosted top pair production and noted a significant excess
above the estimated background of events with two ultra-massive
boosted jets. We discuss the interpretation of the measurement
and its fundamental implications. In case new physics is
involved, the most naive contribution is from a new particle
produced with a cross section that is a few times higher than
that of the top quark and a sizable hadronic branching ratio.
We quantify the resulting tension of a possible larger top pair
cross section with the absence of excess found in events with
one massive boosted jet and missing energy. The measured planar
flow distribution shows deviation from CDF's Pythia QCD
prediction at high planarity, while we find a somewhat smaller
deviation when comparing with other Monte Carlo tools. As a
simple toy model, we analyze the case of a light gluino with
R-parity violation and show that it can be made consistent with
the data.
\end{abstract}

\maketitle

\mysection{Introduction}
New physics searches at colliders typically focus on signals
with leptons and/or missing energy. Recently, there has been
some interest in extending the hunt to include particles that
decay only to quarks and gluons (see \eg~\cite{gluino_searches,
leshouches} for some theoretical studies), as was done in an
analysis by CDF~\cite{cdflowpt}. In this analysis the focus was
on supersymmetry (SUSY) with R-parity violation (RPV), where a
light gluino decays to three quarks. However, this results in a
multi-jet signal, which makes it challenging to distinguish
from the QCD background. Indeed, it was found
in~\cite{cdflowpt} that the current sensitivity is far below
the expected signal, thus it is not useful for obtaining a
bound on the parameter space of SUSY (or any alternative theory
which would produce this type of a signal).

Progress has been recently achieved in another CDF study by
restricting the data sample to include only high transverse
momentum ($p_T$) and high mass jets~\cite{cdfboostedqcd,
cdfboostedtops}, thus reducing the QCD background much more
than the signal and \emph{increasing} the sensitivity (as was
anticipated in~\cite{leshouches}). The idea is that the decay
products of a highly boosted massive object would collimate to
a single jet in the detector. While the data is still dominated
by the QCD background, it has much larger discrimination power.
Moreover, it is possible to use various jet substructure
analysis techniques to further improve the efficiency. Applying
this approach enabled to obtain the strongest existing bound on
the cross section for the production of a (high-$p_T$) top
pair, even without relying on substructure analysis.

The CDF study focused on events including two boosted jets
($p_T>400$~GeV for the leading jet) with mass close to the top
mass (130-210~GeV) and pseudorapidity $\eta < 0.7$ (to be
precise, an $\eta$ cut was applied only for the leading jet,
but it was found that the second jet admitted a similarly
bounded $\eta$ value)~\cite{cdfboostedqcd,cdfboostedtops}. The
jet algorithms used are Midpoint and
anti-$k_T$~\cite{Cacciari:2008gp} with $R=1.0$ ($R=0.7$ was
also checked), which were in excellent agreement. As discussed
below, the estimation of the background depends on a parameter
$R_{\rm mass}$ (see Eq.~\eqref{Rmass1}). Using data sample of
5.95~fb$^{-1}$ and assuming $R_{\rm mass}=1\,$, the standard
model (SM) expected number of events is
\beq \label{exp_background}
\begin{split}
\mathrm{QCD}\big|_{R_{\rm mass}\,=\,1}:& \ 13\pm
2.4\;(\mathrm{stat.}) \pm 3.9\; (\mathrm{syst.}) \,,
\\
t \bar t:& \ 3.0 \pm 0.8 \,.
\end{split}
\eeq
The number of observed events was 32~\cite{cdfboostedtops},
which constitutes a deviation of 3.44 standard deviations
($\sigma$) from the  above expectation. In order to translate
this to cross section, we perform the following exercise. The
SM NNLO cross section for $t \bar t$ production with
$p_T>400$~GeV is 4.5~fb~\cite{cdfboostedtops,Kidonakis:2003qe}.
Multiplying this by a branching ratio of 4/9 for hadronic tops,
we get 2~fb, which corresponds to the 3~events reported in
Eq.~\eqref{exp_background}. Thus the difference between the 32
observed events and the mean value of
Eq.~\eqref{exp_background} is translated to a cross section of
\beq \label{excess}
\sigma_{\rm excess} \sim \left(11\pm 3.2\right)\textrm{\,fb}\,.
\eeq
This is the excess found in~\cite{cdfboostedtops} in terms of
hadronic top-equivalent cross section, under the
\emph{assumption} that the signal cannot be accounted for
within the SM. The data can also be used to provide an upper
bound on the all hadronic top pair production cross section,
which is given by 20~fb at 95\% confidence
level~\cite{cdfboostedtops}.

The evaluation of the QCD background in
Eq.~\eqref{exp_background} was done in the following way. The
search was divided into four different regions in terms of the
jet masses. Region~A corresponds to events with two ``light''
jets, with masses in the range of 30-50~GeV. Regions~B and~C
are for one massive jet (130-210~GeV) and one light jet,
depending on which is the leading jet in terms of $p_T$.
Finally, region~D corresponds to two massive jets. There are
three basic assumptions involved: i) all the events in regions
A-C come only from QCD; ii) the actual cross section can be
factorized into the partonic cross section, which only weakly
depends on the masses of the final states, and the jet and soft
functions~\cite{factor_soft}; iii) the masses of the leading
and sub-leading jets are largely uncorrelated variables for QCD
jet production, and the correlation cancels in the ratio
$R_{\rm mass}$ described below. Under these assumptions, we
have
\beq \label{Rmass1}
R_{\rm mass}\equiv \frac{n_B n_C}{n_A n_D}=1\,,
\eeq
where $n_X$ is the number of events in region $X$. One can
therefore estimate the number of QCD events in region D by $n_B
n_C/n_A$. The result of this calculation is the one given in
Eq.~\eqref{exp_background} for QCD. Below we test this
estimation in more detail.

The CDF study~\cite{cdfboostedtops} used another search
channel, including one jet (with $p_T>400$~GeV and mass
130-210~GeV) plus missing energy (with missing $E_T$
significance between 4 and 10~-- see definition
in~\cite{cdfboostedtops}). In the context of $t\bar t$
production, this corresponds to events with one top decaying
hadronically and the other semileptonically. Note that this
type of measurement suffers from a lower signal to background
ratio, since there are large fluctuations in the jet energy
scale, which make the estimation of the missing energy noisy
(see Fig.~10 in~\cite{cdfboostedqcd}, where there are long
tails for both the $t \bar t$ and QCD missing transverse energy
significance distributions). The total number of events
observed in both channels is 58, the estimated QCD background
(for $R_{\rm mass}=1$) is $44\pm 8.4\;(\mathrm{stat.}) \pm 13\;
(\mathrm{syst.})$ and the $t \bar t$ background is 4.9. This
leads to an upper bound of 40~fb at 95\% confidence level on
the $t \bar t$ production cross section for top quark
$p_T>400$~GeV.

Another result given in~\cite{cdfboostedqcd} is the planar flow
(Pf) distribution~\cite{substructure,topjets} (see
also~\cite{Thaler:2008ju}). This jet substructure variable
distinguishes between a linear deposition of the energy inside
the jet, favored by QCD processes (giving values close to 0 for
Pf), and a planar one (that is, Pf close to 1), produced by the
3-body decay of a top quark. The plot given
in~\cite{cdfboostedqcd} shows that in the data there are more
events with high Pf values than predicted for QCD alone.

\medskip

\mysection{Model Independent Interpretation}
The excess of events with two ultra-massive boosted jets hints
for a contribution which is characterized by a mass scale
around the top one. This new source of massive jets should be
produced with a cross section bigger than that of the SM
hadronic $t\bar t$ by a factor of roughly 5 (about 11~fb in the
signal region, as in Eq.~\eqref{excess}, but not more than
20~fb) and a dominant branching ratio for a fully hadronic
decay. In order to have significant acceptance under the search
criteria, the production should be mostly central, that is with
$\eta \lesssim 0.7$ for both jets. Furthermore, if it is due to
the decay of a massive particle, the collimation rate, which is
the fraction of decays where the daughter particles collimate
into a single jet, must be high, \eg~similar to that of the top
($\sim 0.5$~\cite{topjets}).

The simplest explanation of this excess would be an
underestimation of the QCD production strength (no massive
particle involved). As described above, the existence of an
excess was established based on an estimation of the QCD
background in the signal region~D, without relying on Monte
Carlo (MC) simulations. In this estimation, it was assumed that
the dependence of the partonic cross section on the outgoing
particles' virtuality (jet mass) is negligible\footnote{We are
grateful to Steve Ellis who questioned this assumption.}, as
mentioned in assumption ii above. To estimate the significance
of this effect, we calculated the leading order partonic cross
section\footnote{For the parton distribution functions (PDF),
we used the CTEQ5 Mathematica implementation from
\url{http://www.phys.psu.edu/~cteq/}.} for each of the regions
of jets masses A-D (denoted as $\sigma_X$ for region $X$) with
the virtuality of the particle representing each jet mass,
\beq
\sigma_X=\int \! dp_T dy \, 2p_T \! \sum_{ij} \int_{x_{\rm
min}}^1 \! dx_1 \frac{f_i(x_1,Q^2)f_j(x_2,Q^2)
\sigma_{ij}}{x_1s+u-m^2} \,,
\eeq
where $m$ is the mass of the jet whose rapidity is $y\,$, $s$
and $u$ are Mandelstam variable of the $p \bar p$ system,
$\sigma_{ij}$ is the underlying partonic cross section and
$f_i$ is the PDF at momentum fraction $x$ and energy $Q\,$. The
relation between $x_1$ and $x_2$ and their integration range
are determined by the kinematics (see
\eg~\cite{supercollider}). Now the number of events $n_X$ is
proportional to $\sigma_X$ times the jet mass functions (still
neglecting any jet correlations, as mentioned in assumption iii
above). Since only the latter part factorizes, the ratio of
events $n_B n_C/n_A\,$ used to estimate $n_D$ should be
corrected as follows:
\beq
n_D=\frac{n_B n_c}{n_A} \times \frac{\sigma_A
\sigma_D}{\sigma_B \sigma_C}\,.
\eeq
We found that this correction raises the estimated QCD
background by only about 5\% in the given jet mass
window\footnote{We found no significant sensitivity to
interchanging between CTEQ5M and CTEQ5L and to multiplying or
dividing the energy scale by $2^{1/4}$.}. This substantiates
the reliability of the result of~\cite{cdfboostedqcd,
cdfboostedtops}.

One possible caveat in this argument is that assumption iii
above could turn out to be wrong. If there is some mechanism in
QCD which leads to bias towards two massive jets (relative to
the evaluation used in~\cite{cdfboostedqcd, cdfboostedtops}),
then it might be that the excess of events in region D is
simply the consequence of underestimating the QCD contribution.

The relation in Eq.~\eqref{Rmass1} is examined by MC
simulations in~\cite{AFB}. The results from different MC tools
are shown in Table~\ref{RABCD}. From this we learn that: i) the
deviations from $R_{\rm mass}=1$ are small (within the
systematic uncertainties); ii) the matched MC results, which
include (jj+jjj+jjjj) and are expected to better estimate the
QCD jet mass distribution at large masses, are in very good
agreement with each other (even though they tend not to agree
on the individual jet mass distribution~\cite{topjets}), giving
$R^{\rm MC}_{\rm mass}\simeq0.87$.

\begin{table}[htdp]
\begin{center}
\begin{tabular}{||l|c|c||}
\hline \hline  {MC tool} & Matching & $R_{\rm mass}$ \\
\hline \hline  Sherpa   & Yes & $0.88\pm0.03$\\
\hline MadGraph & Yes & $0.86\pm0.04$\\
\hline MadGraph & No  & $0.76\pm0.04$\\
\hline Herwig   & No  & $0.86\pm0.02$\\
\hline \hline
\end{tabular}
\end{center}
\caption{The results for $R_{\rm mass}$ (borrowed
from~\cite{AFB}) from different MC tools: Sherpa
(1.2.3)~\cite{Gleisberg:2008ta} with matching,
MadGraph/MadEvent 4.4.56~\cite{Alwall:2007st} with MLM
matching~\cite{Hoche:2006ph} to the Pythia package
2.1.4~\cite{Sjostrand:2006za}, MadGraph/MadEvent with no
matching and Herwig 6.520~\cite{Corcella:2002jc} with no
matching. The PDF set used was CTEQ6M~\cite{Pumplin:2002vw},
and FastJet 2.4.2~\cite{Cacciari:2005hq} with anti-k$_t$
algorithm~\cite{Cacciari:2008gp} ($\Delta R=1$) was used for
jet clustering. Quoted errors are statistical only.}
\label{RABCD}
\end{table}%

The other possible explanation would be that the excess is
related to non-SM production of boosted top pairs. A relevant
aspect of the CDF data is that no excess was found compared to
the SM in the channel with one jet plus missing energy
described above. However, this channel suffers from larger
uncertainties, as already mentioned.

In the following exercise we estimate the tension in case the
hadronic excess is completely accounted for by tops. Adding 16
hadronic top events, the expected semileptonic sample (since
including $\tau$'s the ratio is the same) would be
\beq
31+1.9+16 \times(1.9/3)\approx43 \,,
\eeq
where 31 is the expected number of QCD events (estimated as
before using the ratio $n_Bn_c/n_A$), 1.9 is the number of
expected hadronic-semileptonic top events, and thus $(1.9/3)$
is the ratio of acceptance of this sample to the fully hadronic
one, based on the estimation in~\cite{cdfboostedtops}. This
constitutes an excess of 17 compared to the observed 26
events~\cite{cdfboostedtops}. The statistical uncertainty
involved is 8.1 events, while the systematics from the jet
energy scale and jet mass measurements is 30\% of the original
31 expected events. These are combined to a standard deviation
of 12 events, which means that the tension with the
semileptonic sample is at the level of $17/12 \cong
1.4\sigma\,$. Thus we conclude that while a pure top excess is
not perfectly consistent with the data, it is far from being
disfavored.

Further motivation for an excess of boosted tops originates
from the possible relation with the measurement of
forward-backward asymmetry in $t \bar t$
production~\cite{topafb} and specifically the large deviation
recently observed by CDF at high invariant
masses~\cite{Aaltonen:2011kc}. This issue is investigated in
detail in~\cite{AFB,Delaunay:2011gv}.

Finally, it is possible that the data hints for a presence of
new massive particles with a large production cross section and
hadronic final states. Standard hadronic top searches include
b-tagging as a necessary condition. Since these show good
agreement with the SM prediction~\cite{Wagner:2010wd}, the
existence of a new particle which decays to a bottom is
probably disfavored, unless this state would only be produced
with a high boost, where these searches would fail~\cite{KKG}.

Regarding the planar flow distribution, it is interesting to
note that a sizable excess for $\mathrm{Pf}>0.4$, relative to
the Pythia prediction, was found in~\cite{cdfboostedqcd}. This
might motivate a search for particles with 3-body (or higher)
decays effectively (for this purpose, the top's decay is
considered as 3-body).

\begin{figure}[htb]
\begin{center}
\includegraphics[width=.5\textwidth]{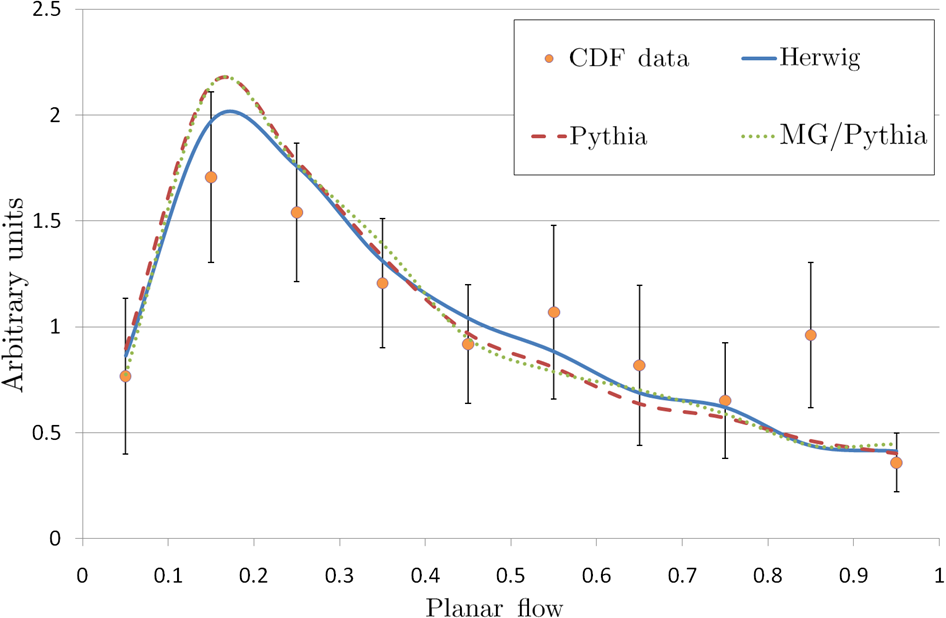}
\caption{QCD planar flow distribution (normalized to unit area) calculated by different
MC tools compared to the CDF data with the anti-$k_T$ jet algorithm
(R=1.0)~\cite{cdfboostedqcd}. The data is represented by
orange circles with error bars, while the solid blue, dashed red and dotted green lines correspond
to Herwig, Pythia and MadGraph with Pythia including MLM matching, respectively.}
\label{qcdpf}
\end{center}
\end{figure}

In order to investigate this issue, we used different MC tools
to estimate the QCD Pf distribution in the relevant search
window. 
The first is Herwig 6.520 with the PDF set CTEQ6L. The second
is Madgraph/MadEvent 4.4.51 with the Pythia 2.1.4 package and
the same PDF set, with and without MLM matching. We also used
Pythia 6.4 by itself. All MCs were interfaced to FASTJET 2.4.2
for jet clustering. The cuts used are the same as in the CDF
study (excluding the $\eta$ cut, which was found to have a
negligible effect). The result is shown in Fig.~\ref{qcdpf},
together with the recent CDF data. It is evident that the three
simulations that we use exhibit good agreement with each other,
and furthermore that their resulting distributions are closer
to the data than the Pythia one in~\cite{cdfboostedqcd}. Note
also that reasonable agreement was found between the
predictions of MadGraph/Pythia and Sherpa in~\cite{topjets}.

\begin{figure}[htb]
\begin{center}
\includegraphics[width=.5\textwidth]{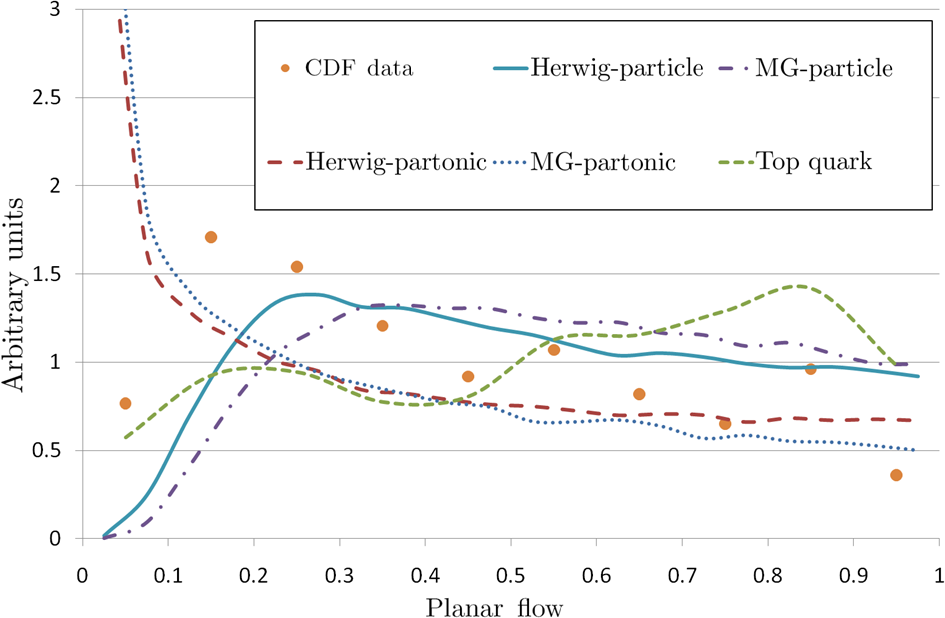}
\caption{Planar flow distribution of an RPV gluino decay (normalized to unit area)
calculated by different MC tools compared to the CDF data with the anti-$k_T$ jet
algorithm (R=1.0)~\cite{cdfboostedqcd}. The data is represented by orange circles.
The solid light blue (dashed red) and dashed-dotted purple (dotted blue) lines
correspond to a particle level (partonic level) simulation using Herwig and MadGraph with
Pythia, respectively. The short-dashed green line is for a hadronic top distribution,
borrowed from~\cite{cdfboostedqcd}.}
\label{gluinopf}
\end{center}
\end{figure}

We further demonstrate that a contribution from particles with
3-body decays favor higher Pf values, such that a proper
combination with the QCD prediction can yield a better
agreement with the data. In Fig.~\ref{gluinopf} we show the
distribution generated by an RPV light gluino (see below),
separating between runs that include only a partonic decay to
three quarks and runs with showering and hadronization (we do
not combine the QCD contribution here). Additionally, the
figure presents the Pf distribution of a hadronic top quark,
borrowed from~\cite{cdfboostedqcd}.

As an exercise, we calculated the Pf distribution of a toy
model where a heavy scalar decays to three massless scalars.
The decay was computed analytically, and the Pf distribution
was obtained by random generation of events admitting the
proper kinematics. It is interesting to mention that the
resulting curve is in perfect agreement with the MG/Pythia
partonic case, while if we add the proper matrix element of the
decay to this ``random'' model, we find perfect agreement with
the Herwig partonic curve.

We note that given the large uncertainties on the data, it does
not seem instructive to make any quantitative comparisons of
the Pf distributions in the two figures. At this stage, both
QCD and 3-body decaying particles provide reasonable fits to
the data. We expect that in the near future, when LHC data is
available, it would be possible to make a distinction between
the different cases~\cite{substructure,topjets,Thaler:2008ju}.

\medskip

\mysection{Toy Model}
In order to demonstrate a toy model that can account for the
observed excess, we consider an RPV gluino in the context of
SUSY, where the rest of the sparticles are decoupled for
simplicity (In principle, there could be interference effects
in gluino production from squarks, but this is highly model
dependent). The gluino decays to three quarks, hence in case
its mass is inside the window used in the search, it would lead
to an excess of events with boosted jets~\cite{leshouches}.

Such a scenario has already received attention in a recent CDF
search~\cite{cdflowpt}, considering only a non-boosted region
with conventional reconstruction. This study focused on signals
of six jets, and employed sophisticated techniques for reducing
the background, such as three-jet correlations and vertex
position tracking. Yet it turned out to be practically
insensitive to a possible gluino contribution.

Another interesting recent work~\cite{leshouches} adopted a
similar approach to that of~\cite{cdfboostedqcd,cdfboostedtops}
in search of an RPV gluino at the Tevatron, though it was based
only on MC simulations rather than real data. It required two
boosted jets ($p_T>350$~Gev) with masses close to each other
and further applied a certain jet substructure cut. This
approach was found to be quite sensitive to a gluino signal.

\begin{table*}[bth]
\begin{center}
\begin{tabular}{||l||c|c|c|c|c||}
\hline
\hline	         				& Acceptance		& Acceptance		  &Cross section [fb] 	&Cross section [fb]    			& Cross section [fb]\\
		Particle      				&  Herwig	 		& MG/Pythia		  &Herwig				&MG/Pythia					&  MG/Pythia\\
                                &                       &no matching        &                    &no matching                & with matching\\
\hline
\hline	Gluino $m_{\tilde{g}}=130$ GeV	& 0.43		    	& 0.49			&15					& 17					& 18\\
\hline	Gluino $m_{\tilde{g}}=150$ GeV	& 0.52			& 0.50			&13					& 14					& 15\\
\hline	Gluino $m_{\tilde{g}}=170$ GeV	& 0.49			& 0.48			&11					& 12					& 12\\
\hline	Hadronic top quark pair 		& 0.47			& 0.46			&1.6				& 1.7					& 1.8\\
\hline \hline
\end{tabular}
\caption{The gluino cross section and acceptance for masses of
130, 150 and 170~GeV, computed by both Herwig and MadGraph with
Pythia. For the latter we also add a calculation including an
extra jet with MLM matching. As a comparison, we present the
hadronic top cross section and acceptance.} \label{crosssec}
\end{center}
\end{table*}

We estimate the gluino signal as a function of its mass using
both Herwig and MadGraph/MadEvent with Pythia. The results are
presented in Table~\ref{crosssec} (note that there is some
difference between the two MC tools, yet it is evident that the
ratio of these cross sections to that of top pair production is
constant). Also shown in the table is the acceptance, which is
the percentage of events that pass all the cuts out of the
overall sample of one boosted jet from the corresponding
particle. It is evident that the cross section is indeed in the
ballpark of the observed excess\footnote{Very recently a new
lower bound of 144~GeV for the gluino mass
appeared~\cite{Aaltonen:2011sg}, thus excluding the first line
in Table~\ref{crosssec}.}. Since we do not try to provide a
precise fit of the signal, NLO corrections are not expected to
change this statement (and in any case they should be small
because of the strong $p_T$ cut~-- see \eg~Figure~9
in~\cite{Ahrens:2010zv}). Moreover, it is interesting that the
gluino contribution enhances the large Pf region of the
distribution, as shown in Fig.~\ref{gluinopf}.

As an outlook to the near future, we point that the LHC should
be able to test whether indeed there is a deviation from the SM
in this type of signal, possibly even with only
$\mathcal{O}(1)$~fb$^{-1}$. It would thus be interesting to
adapt the search of ultra-massive highly-boosted jets to the
LHC.

\mysection{Acknowledgments}
We thank Johan Alwall, Jon Butterworth, Benjamin Fuks, Seung
Lee, Fabrizio Margaroli, Pekka Sinervo and Jay Wacker for
useful discussions. GP is the Shlomo and Michla Tomarin career
development chair and supported by the Israel Science
Foundation (grant \#1087/09), EU-FP7 Marie Curie, IRG
fellowship, Minerva and G.I.F., the German-Israeli Foundations,
and the Peter \& Patricia Gruber Award.


\end{document}